\documentclass[11pt]{article}
\usepackage{verbatim}
\usepackage{geometry, latexsym}
\usepackage[parfill]{parskip}
\usepackage{amssymb,amsmath}
\usepackage{epstopdf}
\usepackage{graphicx}
\usepackage{setspace}
\usepackage{sidecap}
\usepackage{amsthm}
\usepackage{enumerate}
\usepackage{multirow}
\usepackage{graphicx}
\usepackage{caption}
\usepackage{url}

\usepackage[labelformat=simple]{subcaption}

\usepackage{tikz}
\usetikzlibrary{shapes,arrows}

\usepackage{caption}

\tikzstyle{decision} = [diamond, draw, text width=4.5em, text badly centered, node distance=3cm, inner sep=0pt]
\tikzstyle{block} = [rectangle, draw, text width=5em, text centered, rounded corners, minimum height=4em]

\title{Challenges in Modeling Complexity of Neglected Tropical Diseases: \\ Assessment of Visceral Leishmaniasis Dynamics in Resource Limited Settings}

\author{Swati DebRoy$^1$ , 
Olivia Prosper$^2$, 
Austin Mishoe$^1$,
Anuj Mubayi$^{3}$\footnote{Corresponding author, e-mail: amubayi@asu.edu}\\
\\\footnotesize$^{1}$Department of Mathematics and Computational Science,
University of South Carolina-Beaufort, Beaufort, SC, USA\\ 
\footnotesize$^{2}$Department of Mathematics, University of Kentucky, Lexington, KY, USA\\
\footnotesize$^{3}$Simon A. Levin Mathematical, Computational, and Modeling Sciences Center, Arizona State University, Tempe, AZ, USA}

\begin{document}

\maketitle

\begin{abstract}

Neglected tropical diseases (NTD), particularly vector-borne diseases (VBD), account for a large proportion of the global disease burden, and their control faces several challenges including diminishing human and financial resources for those distressed from such diseases. Visceral Leishmaniasis (VL), the second-largest parasitic killer in the world (after malaria) affects poor populations in endemic countries and causes considerable cost to the affected individuals and their society. Mathematical models can serve as a critical tool for understanding the driving mechanisms of a NTD such as VL. 
The WHO promotes integrated control programs for VL but this policy is not well supported by systematic quantitative and dynamic evidence and so potential benefits of the policy are limited. Moreover, mathematical models can be readily developed and used to understand the functioning of the VL system cheaply and systematically.  
The focus of this research is three-fold: (i) to identify non-traditional but critical mechanisms for ongoing VL transmission in resource limited regions, (ii) to review mathematical models used for other infectious diseases that have the potential to capture identified factors of VL, and (iii) to suggest novel quantitative models for understanding VL dynamics and for evaluating control programs in such frameworks for achieveing VL elimination goals. 

\end{abstract}

\section{Introduction}\label{Intro}

Visceral Leishmaniasis (VL) is a neglected vector-borne infectious disease that is transmitted to humans by infected sandflies and is the second-largest parasitic killer in the world after malaria \cite{chappuis2007visceral,WHOFactSheet}. If left untreated, most cases result in death within two to three years of clinical manifestation. Most of the new cases (approximately 90\%) occur in Bangladesh, Brazil, Ethiopia, India, Nepal, South Sudan, and Sudan. VL is identified as a Neglected Tropical Disease by the WHO because it is endemic in several poverty stricken regions of the world, although preventive measures and successful treatment is common in most developed countries. Many people living in these impoverished regions are daily-wage workers, for whom infection with a disease like VL restricts the bread-winners' ability to provide livelihood for their families. Moreover, the cost of treatment and duration of stunted income pushes them into a vicious cycle of further hardship and irrecoverable financial deprivation. Although local government authorities and the WHO have devised several control programs to lower the burden of VL in these regions, the VL endemicity always creeps back after a brief period of relief. This ineffectiveness has been attributed to several factors, including severe under-reporting of cases and death due to VL, lack of clarity in the etiology of the disease, and limited estimation of reservoirs of the infection. Thus, the intensity and extent of the control programs were in conflict with the magnitude of the true VL burden. With limited resources available in many of the affected countries, mathematical modeling can help shed light on several of these challenges cheaply (including identifying cost-effective driving mechanisms), as it has done for other infectious diseases like malaria. Hence, immediate attention from the modeling community is in dire need.

In the past, the WHO has set several elimination target dates for VL, which could not be achieved in the Indian subcontinent. The primary reason for this shortcoming may be the  ineffective implementation of policies in the face of a developing country's infrastructure. Mathematical modeling approaches in conjunction with model guided additional field research in India could be a turning point for achieving optimal program implementation and may help to (1) quantify the ``true" burden of VL in Bihar where it has proven to be particularly difficult to eliminate, (2) investigate the potential mechanisms for the spread of the Leishmania parasite, and (3) suggest optimal vector control programs that may help in achieving the WHO goal of elimination of VL by the year 2017 \cite{le2016feasibility}. 
The VL elimination program target is to reduce the annual incidence to less than 1 per 10,000 at the district or sub district level in South Asia by 2017 (WHO). Currently, the incidence is around 20 cases per 10,000 \cite{mondal2009visceral, dhillon2008national,chowdhury2016implication}.
Understanding the mechanisms driving the transmission dynamics of VL may require the study of several factors, including complex interactions of multiple reservoir hosts, environment-dependent vector dynamics, changes in political and public health policy, spread of resistance to insecticides and drugs, and short and long term human migration patterns.


Since Sir Ronald Ross' first paper using a mathematical model to study the transmission dynamics of malaria in 1906, there have been many modeling studies focusing primarily on infectious diseases; however, more studies are needed on neglected tropical diseases such as VL. One of the aims of this study is to suggest mathematical modeling approaches 
for capturing identified 
regional issues that may be critical in better evaluating control programs in resource limited settings, thereby, assisting in the development of cost effective elimination strategies. We discuss specific features of VL (including treatment availability, living conditions, effect of social status, and implementation cost of control programs) that should be incorporated into quantitative methods that can be effectively used to analyze control strategies for the disease. Carefully established and directly relevant mathematical models are urgently needed  for VL control in an effort to develop a suitable tool to truly capture the complex dynamics of the disease within the given natural or man-made environments and to achieve elimination goals. The assessment of a tropical disease risk must be interpreted on the basis of local environment conditions, the effects of socioeconomic development, and its capability to effectively sustain control programs.

%
%
%

\section{Challenges for Leishmaniasis in Resource-Limited Regions}\label{Risks}

\begin{figure}[htbp] 
   \centering
   \includegraphics[width=5in]{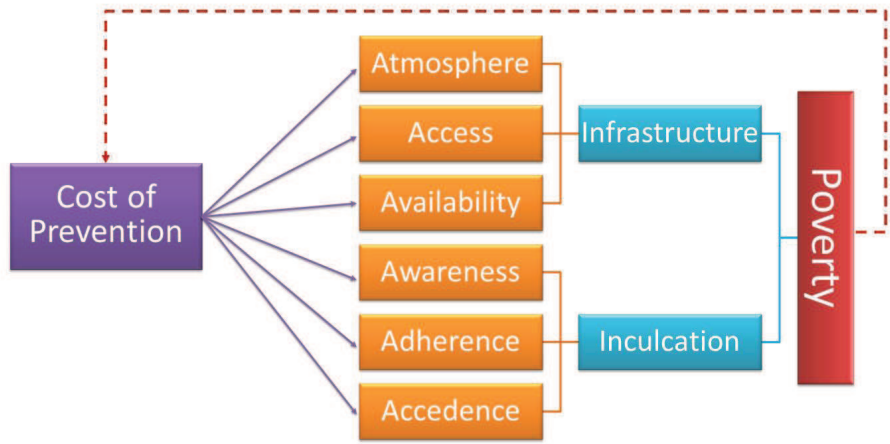} 
   \caption{A viscious cycle of socio-economic challenges and difficulty to access disease interventions}
   \label{fig:cycle}
\end{figure}

When considering the socio-economic challenges of a neglected tropical disease at the  grassroots level, the depth and complexity appears overwhelmingly varied. So, to better comprehend the nature of obstacles, we classified some of the key issues in to the six categories viz., atmosphere, access, availability, awareness, adherence and accedence (6 A-s), which in turn can be traced back to lack of  either inculcation or infrastructure. Each of these variables ultimately stems from the sheer poverty and its viscious cycle with diseases in the affected region (Figure \ref{fig:cycle}). In this review we focus on the state of Bihar, which hosts 90\% of India's VL cases (WHO), and its neighboring countries of Nepal and Bangladesh where the disease dynamics is similar. The categories are briefly described below.

\textbf{Atmosphere:} In the state of Bihar, the worst affected areas are in remote agricultural villages. Studies have revealed several living conditions positively correlated with higher prevalence of leishmaniasis. In two independent studies, factors like mud plastered houses, vegetation and bamboo near the house, and granary inside the house were found to significantly contribute to leishmaniasis \cite{dhiman1991epidemiology, ranjan2005risk}. Other systematic studies in the subcontinent with similar geographical settings (Uttar Pradesh, India; West Bengal, India; Terai, Nepal; Mymensingh, Bangladesh) have found that living conditions with cracked mud walls, damp floors, and close proximity to a water body are risk factors for leishmaniasis. Also, a high density of occupants in a household with more than three people per room were found to increase transmission \cite{bern2007epidemiology, nandy1987leishmanin, barnett2005virgin, bern2005risk, bern2000factors, ranjan2005risk, schenkel2006visceral, saha2009visceral,  rukunuzzaman2008epidemiological}. Models that have incorporated such features (house types, household size etc.) exists for other diseases \cite{yong2015agent,kasaie2014timing, perez2009agent, akhtar2007chain, noiva2014susceptible, cohen2001modeling}. 


It is well known that sand-fly bites thrive during the warmer months (March - June, October), and late in the evening \cite{dinesh2001seasonal}. The role of climatic factors on transmission dynamics of vector-borne diseases has been thoroughly studied in the literature \cite{artzy2010transmission, parham2010modeling}. The hot Indian summer in combination with lack of electricity often lead people to sleep outdoors, which increases the number of sand-fly bites and hence the risk of contracting leishmaniasis \cite{barnett2005virgin, bern2000factors}. Understandably, proper use of bed-nets have been found to have a protective effect on people across several studies \cite{bern2007epidemiology, barnett2005virgin, bern2005risk,  schenkel2006visceral, saha2009visceral, rukunuzzaman2008epidemiological} and sleeping on a cot (versus on the floor) also demonstrated a protective effect. 
Models have shown that proximity to domesticated animals was found to play a complex role in containing, spreading and serving as a possible reservoir of the parasite \cite{gorahava2015optimizing}. For example, some studies found that proximity to livestock provided a protective effect against leishmaniasis \cite{bern2005risk,  bern2000factors}, whereas in Uttar Pradesh, India, the risk of leishmaniasis was found to increase with increased numbers of cattle in the vicinity of a household \cite{barnett2005virgin}.

\textbf{Access:} Currently, therapuetic interventions for Kala-azar (Indian VL) are significantly subsidised by the ministry of health in India (National Vector Bourne Disease Control Program's Kala-azar Elimination Initiative under the Govt. of India). There are 38 District Health Societies (DHS) inside the state  healthcare system (State Health Society) of Bihar. The DHS are further subdivided into a number of block-Primary Health Centers (PHCs); the number of PHCs per DHS varies for each DHS. Again, the PHC consists of Sub-Centers providing health care to a certain number of villages (e.g. the Muzaffarpur district (population 3.7 million) has fourteen Block PHCs while the Kanti Block (population 337,670, Census of India, 2001) has 48 Sub-Centers, each covering a population of roughly eight thousand individuals \cite{Singh2006serious}). PHCs, district hospitals and government medical colleges are the sources of reported cases (National Vector Borne Disease Control Programme, 2009). The private health sector includes not-for-profit and for-profit organizations. For-profit venues include corporations (e.g., private nursing homes), trusts, stand-alone specialist services, pharmacy shops, and self-appointed practitioners. Estimates suggest that 80\% of the outpatients and 57\% of the inpatients are handled in the private sector (The World Bank report, 2001 \cite{peters2001raising}). NGOs usually provide awareness and education programs, carryout research, and provide access to regular health services. Ninety percent of Bihar's population lives in rural areas where less than 1\% of health services are provided by not-for-profit/Non-Government Organizations (NGOs) (The World Bank, 2001 \cite{peters2001raising}). Patients in rural areas travel on average much further for treatment than patients in urban areas. Thus, access to healthcare can be tricky at present and efforts need to be made to encourage the set-up of temporary mobile clinics in harder to reach areas and to encourage people to seek out certified treatment. 
	Bihar is the poorest state in India, where the ``caste" (proxy for social standing) of a person is born into affects almost every aspect of the social conduct he/she receives their entire life. \cite{van2009decomposing}. Martinez \cite{Martinez2012} found that the people of lower caste are consistently being seen by a doctor at a more advanced stage of VL than those of a higher caste. In fact, most VL patients in the disadvantaged caste see a doctor more than eight weeks post symptom onset, which includes a larger spleen and lower hemoglobin level than normal. Thus, efforts need to focus more on the people of lower caste to diminish the disparity in healthcare; only then can planned control measures effectively reduce the overall burden of VL. Models considering underreporting of cases due to treatment of patients by non-reporting private healthcare clinics and patients' healthcare seeking behaviors have been developed \cite{mubayi2010transmission, medley2015health}.

\textbf{Availability:} The WHO recommends the use of a single dose of Amphotericin B as the first line of treatment in the Indian sub-continent \cite{matlashewski2011visceral}. However, daily injections of Pentavalent Antimonials (SSG) for 20-30 days and 15 injections of Amphotericin B every other day are still more widely used in India (NVBDCP). The availability of drugs in a timely manner is dependent on several factors including the affordability of a drug by the government, reasonable forecasting of the quantity of drug required (to avoid shortage as well as waste), proper storage and distribution of the drugs throughout the lengthy route from the manufacturer to the affected people, avoiding cheaper counterfeit drugs and also drug legislation \cite{den2011leishmaniasis}. Mathematical models can capture each of these features \cite{moghadas2008population, katouli2011worst} and study their role in the spread and transmission dynamics of VL.

\begin{figure}[htbp] 
   \centering
   \includegraphics[width=6in]{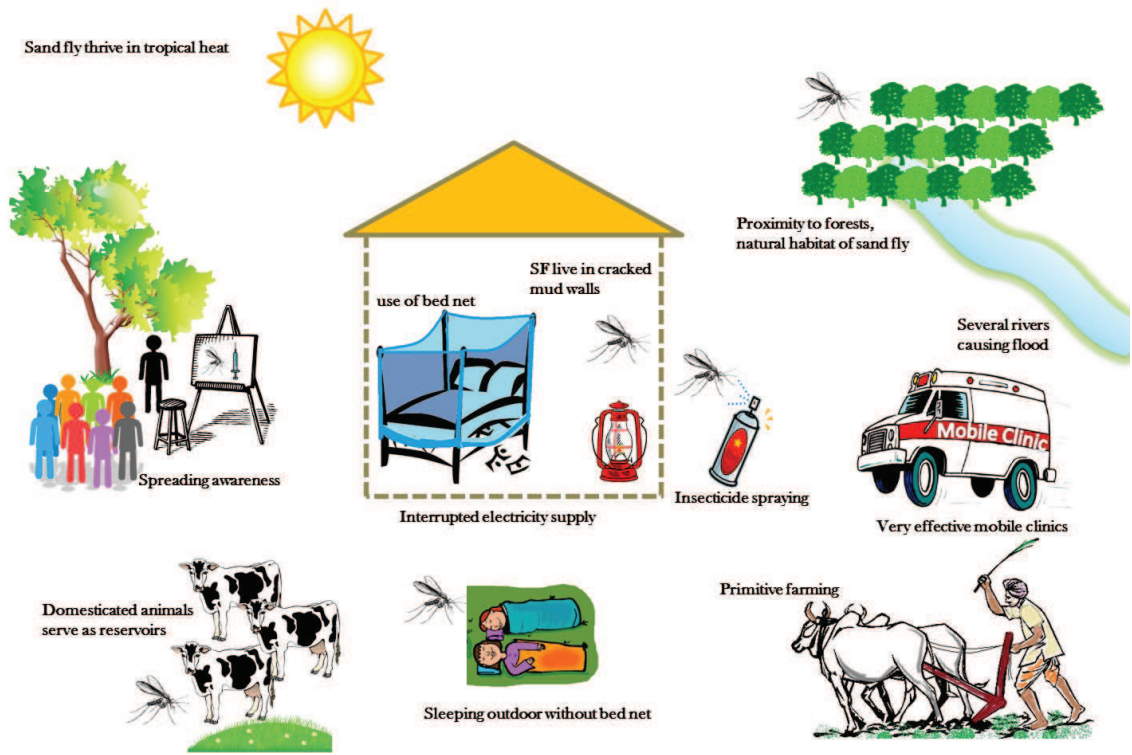} 
   \caption{Cartoon reflecting social aspects on which awareness programs to control spread of VL can be designed to reduce disease burden}
   \label{fig:examples_programs}
\end{figure}

\textbf{ Awareness:} `Awareness' can be defined as knowledge regarding the etiology of the disease which would help local individuals to prevent infection and to look out for VL symptoms and seek medical attention sooner rather than later. Figure \ref{fig:examples_programs} shows some social aspects for which awareness programs may be needed as a prevention for VL. Lack of awareness causes a disease which is curable upon treatment, to end up causing death. Even symptomatic VL-infected people mix in the population freely, thus considerably increasing the chances of transmission. In a study on Nepal (which shares a boarder with Bihar) by Rijal 2006 {\it et al.}, it was found that affected people from the poorest strata of the community preferred to visit a private doctor or local faith healer over public health clinics, leading to higher costs for these individuals \cite{rijal2006economic}. Moreover, debts aquired during this period, in addition to lack of income (the earning adult being ill), creates a major financial abyss which is almost impossible to recover from. Thus, it is not sufficient that the government provides free treatment to the people, it is also necessary to disseminate that information in an effective manner to every strata of people in the affected region. In a study in Brazil, significant awareness was spread in communities through educating school children, who in turn were assigned to discuss interventions mentioned in student's homework assignments with their family members \cite{magalhaes2009dissemination}. Models have studied the role of awareness programs on transmission dynamics of diseases \cite{pittet2006evidence, bhunu2011mathematical}.

\textbf{Adherence:} Non-adherence to treatment is a major factor contributing to the high development of resistance to pentavalent antimonials in the population exposed to VL in the Indian sub-continent \cite{den2011leishmaniasis}, and subsequently resulted in it being discontinued as a first line treatment for VL. 
There are two major factors which contribute to non-adherence in the region: lack of knowledge about the consequences of incomplete treatment leads to patients stopping treatment once the symptoms are relieved, and the financial loss due to reduced days of productivity while on therapy is a major deterrent to continuing treatment. It has been well documented in a 2000-2010 cohort study in Nepal that another disease, Post Kala-azar Dermal Leishmaniasis (PKDL), a sequel and reservoir for VL, was more common in patients who were inadequately treated during VL versus the ones who adhered to the full course of treatment. The overall prevalence of PKDL in SSG treatment was 2.9\%, 0.3\% in supervised and 4.5\% in unsupervised treatment \cite{uranw2011post}.
As observed by Rijal 2006 {\it et al.}, loss of productivity implies no income at all for the poor families \cite{rijal2006economic}. Farmers are unable to attend to their fields, possibly during very important farming phases which results in lowered income for a considerable period of time. Alot of people in the poorest section of society in Bihar are also daily-wage earners and each missed day of work might present dire consequences for the entire family. To survive this period, the family takes out loans from private lenders at high interest, ultimately leading them to further poverty.  
Thus, it is important to inform the people that non-adherence would lead to relapses which are harder to treat; although the burden of missed work is difficult to accept in the immediate context, it is better ultimately for the family. Consequently, providing financial assistance to affected families while their primary earning members are under treatment may improve the outcome of VL control programs, by encouraging greater adherence to treatment. Although there is a need for better models that can capture irregular treatment adherence levels among patients, some simple models exist in the literature \cite{huang2008modeling, smith2006adherence, kalichman2010adherence, fisman2014projected}. 


\textbf{Accedence:} `Accedence' is defined as the acceptance to undergo testing and treatment for PKDL, which can occur in patients who have recovered from VL post-treatment caused by the same pathogen. As reported by Desjeux {\it et al.} in 2013, patients with PKDL serve as a reservoir of VL and it is unlikely that VL can be eradicated without addressing the issue of diagnosing and treating PKDL \cite{desjeux2013report}. In fact, aggressive measures are required to encourage people to consult a doctor in case of any persistent skin lesions, and once diagnosed these should be treated with Amphotericin B, which has been found effective in the Indian sub-continent. 
As reported by Thakur {\it et al.}, attempts were made to fast-track diagnosis and treatment of PKDL in several badly affected districts of Bihar and yielded a very positive outcome \cite{thakur2009newer}. However, in addition to the use of mathematical modeling methods, these efforts need the support of the local governmental authorities to continue and succeed where it is most needed. 

It is evident that aggressive control measures are necessary to address every issue and to ultimately alleviate the burden of VL from Bihar. However, several of the issues arise from limited resources in the affected region and thus the control strategies should be carefully weighed by importance and cost-effectiveness. Moreover, it is not only important to take drastic measures using the one-time funds generated for this purpose, but it is essential that powerful and sustainable changes in the system are established through easy and systematic ways of approaching the difficulties, which can only be attained by identifying critical mechanisms of the system. Mathematical models are one of the best methods to cheaply and systematically identify driving factors.

\section{Review of VL Mathematical Modeling Studies}\label{Modeling}

Despite the incalculable harm and countless challenges leishmaniasis inflicts on populations around the globe, only a handful of publications address the problems from a mathematical modeling perspective.  In fact, a recent review by Rock {\it et al.}, which tabulates all mathematical models of VL, found only twenty-four articles using mathematical models for VL, several of which used the same base model structure \cite{rock2015uniting}.  Of these twenty-four articles, only seven addressed VL in the Indian Subcontinent.  Arguably, one of the greatest modeling challenges is the limited understanding of the leishmania pathogen, the sandfly vector, and how disease manifests in humans.  Dye {\it et al.} \cite{dye1988earthquakes} spear-headed the application of mathematical models to leishmania dynamics.  The authors developed a simple discrete-time model with {\it Susceptible}, {\it Infected}, and {\it Resistant} humans to study the mechanism behind inter-epidemic periods observed in VL cases between 1875 and 1950 in Assam, India.  Counter to the existing theory of the time, the model demonstrated that the observed inter-epidemic patterns could be explained by intrinsic factors in leishmania transmission.  This modeling effort also stressed the significance of PKDL and sub-clinical infections in determining whether a region will have endemic or epidemic leishmaniasis.  A few years later, Hasibeder {\it et al.} \cite{hasibeder1992mathematical} published a compartmental delay-differential equation model for canine leishmaniasis.  This model accounts for two types of dogs: those that will develop symptoms, and those that will remain asymptomatic, following infection by a sandfly.  The model also explicitly describes the infection dynamics of the sandfly vector and considers a fixed delay representing the extrinsic incubation period.  The authors take a heuristic approach to derive a formula for the basic reproduction number $R_0$, the number of secondary sandfly infections resulting from a single infected sandfly, in an otherwise fully susceptible population.  Although this model addresses two important aspects of the natural history of the disease that may be extended to human VL, namely the asymptomatic human and infected vector populations, the model does not consider the asymptomatic population to be an infectious reservoir, assumes constant human and vector population sizes, and omits the effects of seasonality.  The model does, however, introduce heterogeneous biting, determined by a dog's ``occupation".  The mathematics developed in \cite{hasibeder1992mathematical} was applied to age-structured serological data for the dog population in Gozo, Malta in \cite{dye_epidemiology_1992}, and provided estimates for $R_0$.  This modeling work was extended in \cite{dye1996logic} to include zoonotic transmission, that is, humans, dogs, and sandflies, were explicitly modeled.  Dye conducted a sensitivity analysis to determine which of three control measures would be most effective in decreasing disease prevalence.  Their results suggest vector control and vaccination of the human or dog population would be more effective than treating or killing infected dogs.


More recently, Stauch {\it et al.} developed a more comprehensive model of VL for the Indian subcontinent \cite{stauch2011visceral}, and later extended it to include drug-resistant and drug sensitive {\it L. donovani} parasites, with a focus on Bihar, India \cite{stauch2012treatment}.  The model proposed in \cite{stauch2011visceral} extended the basic SIR model structure for the human population by segmenting the infected stage into five distinct stages according to an individual's infection status determined by the results of three diagnostic markers.  These diagnostic markers were (1) PCR, the earliest infection marker, (2) DAT, which measures antibody response, and (3) LST, which can detect cellular immunity.  The model also includes treatment of symptomatic VL cases, treatment failure, relapse characterized as PKDL, PKDL treatment, and HIV-co-infection (described in their Supplementary materials).  The sandfly population is modeled using an SEI ({\it Susceptible-Latent-Infectious}) model.  Treatment of VL is divided into first and second-line treatment, and treatment-induced mortality caused by drug-toxicity is considered.  The model was fitted to data from the KalaNet trial using Maximum Likelihood.  The authors explored several intervention strategies, including treatment, active case detection, and vector control.  Although the authors warn that their model assumes homogeneous transmission, ignoring possible clustering of cases within affected households, their modeling approach and parameter estimation argues that the large asymptomatic reservoir precludes the ability for a treatment-only control program to attain the desired target of less than 1 case per 10,000 individuals annually.  Vector-based control is much more promising, but the authors estimate it can only reasonably reduce VL incidence to 18.8 cases per 10,000.  Consequently, the authors emphasize the need for active case detection, effective treatment of PKDL, and vector control to achieve VL elimination.  

Based on the model in \cite{stauch2011visceral}, and following up on their finding that treatment of VL does little to reduce transmisson, Stauch {\it et al.} investigated the uncertainty in their parameter estimates and explored the efficacy of different vector-based control measures in \cite{stauch2014model}.  They estimated that $R_0$ for VL is approximately 4.71 in India and Nepal, and that reducing the sandfly population by 79\% via reduction of breeding sites, or reducing the sandfly population by 67\% through increasing sandfly mortality, are both sufficient to eliminate the {\it L. donovani} parasite in the human population.  The authors argue that recent evaluations of IRS (indoor residual spraying) efficacy suggest that elimination should be possible, with the caveat that the situation may change if insecticide resistance emerges.  However, vector management using LLIN's (long-lasting insecticide-treated nets) or EVM (environmental management) would not be sufficient to achieve elimination.  The authors emphasize the need to study infection rates, the parasite dynamics in both the human and vector population, animals serving as alternate hosts or potentially infection reservoirs, and the contribution of the asymptomatic population.  Furthermore, Stauch {\it et al.} suggest extensions of the deterministic modeling framework to include heterogeneity and seasonality.

In \cite{stauch2012treatment}, Stauch {\it et al.} extended the model in \cite{stauch2011visceral} to include both resistant and sensitive parasites.  The authors considered five mechanisms by which the fitness of the resistant strain may differ from the sensitive strain: (1) increase probability of symptoms, (2) increase parasite load, (3) increase infectivity of asymptomatic humans, (4) increase transmission from symptomatic and asymptomatic host to vector, (5) increase transmission from vector to host.  Simulations of this extended model indicate that a treatment failure rate over 60\% is required to explain observations in Bihar.  Furthermore, observations in Bihar cannot be explained without assuming an increase in fitness in resistant parasites.  The authors explain that it is more likely that the necessary additional fitness is transmission-related rather than disease-related.  Unfortunately, their results also suggest that once a more fit resistant parasite has been introduced, that parasite will eventually exclude the sensitive parasite, even in the absence of treatment.

The work of Mubayi {\it et al.} \cite{mubayi_transmission_2010} is the first to use a rigorous, and dynamic model to estimate underreporting of VL cases at the district level in Bihar, India.  The authors designed a staged-progression model, composed of a system of nonlinear, coupled, ordinary differential equations.  The stage-progression model exploits the fact that the sum of $n$ independent exponential distributions with rate parameter $\lambda$ is a gamma distribution with shape parameter $n$ and scale parameter $1/\lambda$ ($\Gamma(n,1/\lambda)$), and captures the observed variability in the incubation period, infectious period, and treatment duration.  Furthermore, the authors address the differences between public and private health care facilities in their treatment and reporting practices by assuming a fraction of infected individuals $p$ seek treatment in public health care facilities, and the remaining proportion seek treatment in private clinics.  Building an empirical distribution for this parameter $p$ and deriving a relationship between model parameters and monthly reported incidence data allowed the authors to estimate the degree of underreporting for each district for the years 2003 and 2005.  This model analysis informed by incidence data revealed that districts previously designated as low-risk areas for VL are actually likely to be high-risk: the true burden masked by underreporting.

ELmojtaba {\it et al.} presented a more classical approach to modeling VL, with a focus on Sudan, in \cite{elmojtaba2010mathematical, elmojtaba2010modelling, elmojtaba2013vaccination}.  Because leishmaniasis in the Sudan is zoonotic, the authors included the dynamics of an animal reservoir in \cite{elmojtaba2010mathematical}.  This baseline model is extended in \cite{elmojtaba2010modelling} to address parasite diversity, and in \cite{elmojtaba2013vaccination} to explore the potential impact of mass vaccination in the presence of immigration.  

All of these modeling efforts have either contributed to our understanding of VL or highlighted the need for better data to construct and validate future models of VL.  However, there are currently no models, to the best of our knowledge, that attempt to link socio-economic factors, like the 6 A's discussed in Section \ref{Risks}, to disease transmission. 


\section{Bridging Socio-economic Factors and Mathematical Models}

In this section, we provide some examples of published mathematical modeling studies where researchers have attempted to incorporate some of the factors mentioned above and studied their role in the transmission dynamics of infectious and physiological diseases. Existing models of visceral leishmaniasis, though limited in number, have incorporated some of the biological complexity, contributing to a more developed understanding of the disease.  However, to formulate applicable control measure recommendations with cost-estimation, models which can simultaneously incorporate the discussed risk-factors explicitly would be necessary. Many of the techniques to incorporate these factors individually can be drawn from the literature regarding heavily studied diseases like HIV, malaria, and tuberculosis.

In a simplistic mathematical model we can incorporate several risk factors associated with ambience implicitly through the interpretation and calculation of the model parameters. For example, the transmission parameter can be considered as a product of the predominant type of housing, density of vegetation around houses, number of domesticated animals, and number of inhabitants in a house. 


Lipsitch {\it et al.} addressed adherence to treatment and its role in promoting drug resistance in a mathematical model for tuberculosis (TB) in the presence of bacterial heterogeneity \cite{lipsitch1998population}.  To model non-compliance to drug therapy, the authors assumed that non-compliant patients adhere to the treatment regimen when bacterial loads are above a certain threshold, and will halt treatment when bacterial loads fall below this threshold, that is,
\[ Adherence\_level(t) = \begin{cases} 
       \frac{\alpha B(t)}{K+B(t)} & if B(t)\geq N_{min} \\
      0 & if B(t)< N_{min} 
   \end{cases}
\]
where $B(t)$ is the bacterial load at time $t$ and $N_{min}$ is the theshold minimum bacterial load under which drug treatment is discontinued. Simulation and analysis of their stochastic-deterministic hybrid model illustrated that non-compliance is one mechanism that can give rise to bacteria resistant to one or more drugs in a multi-drug therapy.  Furthermore, the authors noted that the pattern of resistance driven by non-compliance more closely resembled observations of patients on multi-drug therapy, compared with the pattern of resistance promoted by bacterial heterogeneity.  Consequently, the model suggests that non-compliance plays a larger role than heterogeneity of the bacteria population in promoting resistance during multi-drug therapy.  The authors noted that an exception to this pattern occurred in HIV-positive TB patients.  The modeling assumption for non-compliance in this TB model addresses one of the `adherence' concerns discussed in Section \ref{Risks}, namely that patients often stop treatment once symptoms are relieved, suggesting a possible framework in which to study adherence to treatment, treatment failure, and if tied to a population-level model, the spread of drug resistant parasites in VL-endemic regions.  

\begin{figure}[htbp] 
   \centering
   \includegraphics[width=6in]{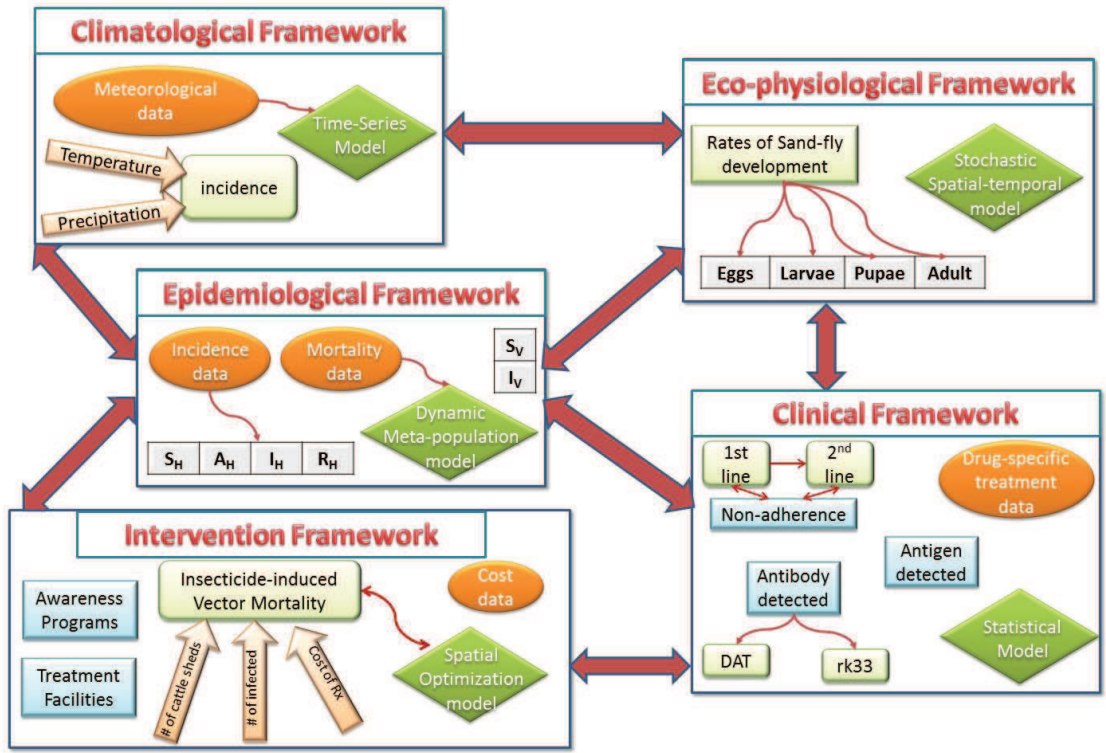} 
   \caption{General categories of modeling frameworks for studying dynamics of vector-borne diseases and types of data needed in such frameworks}
   \label{figmathmodel}
\end{figure}

Adherence to treatment is also a concern in diabetes patients, despite the fact that non-adherence increases the likelihood for stroke and other potentially fatal complications.  Mason {\it et al.} \cite{mason2012optimizing} developed a Markov Decision process (MDP) model to study the timing of treatment initiation and drug-adherence in diabetes patients and the role these two factors play in determining a patients' quality-adjusted life years (QALYs).  Consistent with observations of adherence behavior in diabetes patients, the model assumed that a patient's health status does not influence future adherence.  This assumption may be relevant for some VL-endemic regions where non-adherence is a consequence of insufficient inculcation of the risks associated with improper treatment, or a result of the cost of treatment.  The model also assumed, consistent with clinical practice, that if a patient or the patient's physician had not already decided to begin treatment, treatment would be immediately initiated following a non-fatal complication. The reward function, dependent on adherence, included several important factors, including QALY, the cost of treatment and hospitalization, and disutility resulting from treatment side effects.  The authors also developed a cost function, with the goal of optimizing treatment initiation, in the presence of different degrees of non-adherence, and compared the optimal timing for `uncertain adherence' and `predictable adherers'. The authors quantified the benefit of treatment relative to the cost through a reward function $r_t(l,m)$, where  $l$ denoted the patient's health status, and $m$ equaled zero or one, depending on whether the patient was on treatment or not. 
$$r_{t}(l,m)=R(l,m)-C_{t}^{0}-(CF^{S}(l)+CF^{CHD}(l))-mC^{ST}(A)-(C^{S}(l)+C^{CHD}(l)),$$
 for $t=1, \dots,T-1,l\in L,m\in M,$ where
$$R(l,m)=R_0(d^{S}(l))(d^{CHD}(l))(md^{ST}(A)),~ l\in L,~ m\in M$$  describes the reward for 1 quality-adjusted life. The decrement factors $d^{S}, d^{CHD},$ and $d^{ST}$ denote the decrease in quality of life from a stroke (S), a coronary heart disease (CHD) event, or statins initiation (SI), respectively. The costs $C^{O}, C^{ST}, C^{S}$ and $C^{CHD}$, and $C^{FS}$ and $CF^{CHD}$ represent the cost of other health care for diabetes patients, cost of statin treatment, cost of initial hospitalization for stroke and CHD events, and cost of follow-up treatment for stroke and CHD events, respectively. This diabetes model suggested that initiation of treatment should be delayed in individuals predicted to have poor future adherence.  Furthermore, the model predicted that over time, only 25\% of patients will remain adherers for greater than 80\% of the days during the study.  
  
The effect of change in disease dynamics due to behavioral change and educational awareness have been modeled using ordinary differential equations-based models in several studies. 
Hallett 2009 analyzed the effect of behavior change on the course of an HIV epidemic \cite{hallett2009assessing}. In their dynamic model, the behavior is incorporated by considering parameters such as mean rate of partner change and condom use in casual relationship as a step-function depending on time at which the change in behavior occurred and time it takes to reach a new value. In Mushayabasa 2012, an ordinary differential equation model was used to quantify the role of an educational campaign in controlling Hepatitis C among women in prison \cite{mushayabasa2012assessing}. Here, the effect of this campaign is reflected as an efficacy factor in conjunction to the parameters which represent the sharing of contaminated needles or syringes among the susceptible and exposed classes. 

Ideas for VL modeling should also draw from modeling techniques used in economics in the context of social sciences to effectively optimize the cost and strategy in a resource-limited region. Fenichel {\it et al.} used an economic behavioral model in conjunction with the classical Susceptible-Infected-Recovered (SIR) model that explicitly models adaptive contact behavior \cite{fenichel2011adaptive}.  The authors construct a utility function, an index which describes an individual's well-being.  The framework assumes that individuals make choices that maximize their utility.  These decisions then impact disease risk, creating a feedback loop between disease risk and decisions made based on perceived disease risk. The authors demonstrate that fitting the classical SIR model to data generated by their new framework results in erroneous estimates of epidemiological parameters, because of its inability to jointly estimate behavioral and biological parameters.  
See Perrings {\it et al.} for a thorough review on the growing topic of economic epidemiology \cite{perrings2014merging}.

A study by Gorahava {\it et al.} 
develops a dynamic optimization model to suggest novel ways of allocating insecticides in the districts based on sizes of both human and cattle populations while considering limited financial and resource constraints \cite{gorahava2015optimizing}. The model maximizes number of sandflies killed by insecticides intervention and minimizes number of human cases while identifying optimal number of houses and cattle sheds to be sprayed for a given budget. 

The results of the above models addressing adherence to treatment, adaptive human behavior, and resource constraints emphasize the need to bridge the gap between socio-economic factors and existing VL modeling frameworks.  The 6 A's should be systematically incorporated into VL model frameworks to assess the sensitivity of model dynamics to these six socio-economic factors.  Failing to address factors that result in significant changes in disease dynamics may result in models that do not effectively inform public health policy.  Likewise, models that do not acknowledge resource constraints may lead to infeasible control policies.

\section{Discussion}\label{Dis}

Leishmaniasis continues to spread in every continent on the planet except Australia and Antartica, and VL is most common in the poorest regions of modern human civilization. The WHO has identified leishmaniasis as a neglected tropical disease owing to its endemicity in the under-developed tropical regions of the world, in spite of available treatment options in first world countries. The reason for the ongoing spread and failure to control lies mainly in underreporting of the disease burden \cite{mubayi2010transmission}, poor infrastructure, lack of awareness, poverty and inadequate control measures. In this review we have presented some of the less highlighted, but nonetheless very important, factors which are key in one of the VL affected neglected regions, the Indian state of Bihar, and possibly in several other impoverished regions; these factors may also play a critical role in the transmission of other NTDs \cite{mubayi2010transmission}.

Mathematical models have been used to understand disease dynamics in other parasitic infections and recommend control measures under different constraints. Thus, we propose mathematical modeling as a cheap and effective tool to devise meaningful control measures that will make the next WHO leishmaniasis elimination goal a reality. Most of the mathematical modeling research on leishmaniasis has focused on the development and choice of drugs and co-infection with other diseases \cite{hussaini2016mathematical,rock2015uniting,stauch2012treatment}. Although the importance of these topics cannot be overlooked, more attention needs to be focused on socio-economic issues leading to lack of infrastructure, inculcating awareness, and promoting healthier practices. In this article we identified key issues relating to these factors as observed and published in the literature.  We also reviewed current mathematical models used for leishmaniasis and discussed some ways of explicitly incorporating these socio-economic issues into mathematical models. 
In light of the major financial constraints in the affected regions, a hybrid dynamic optimization model (an example framework is shown in Figure \ref{figmathmodel}) will be necessary, which can calculate monetary (cost of interventions) and non-monterary (mortality and morbidity) factors related to VL for the specific region taking into consideration the socio-economic drawbacks. The building of such a model will require detailed quantification of every aspect of life in the regions, including non-tangible issues. Moreover, the execution of this model will require extensive sets of data on these varied aspects, which can be challenging considering the current dearth of data. However, in absence of data, projections can be made for different scenarios to elucidate an understanding of the magnitude of the problem, and to estimate the relative importance of different socio-economic factors to accurately predicting disease dynamics and informing effective public health policies.


\end{document}